\documentstyle[12pt,twoside,fleqn,espcrc1]{article}

\newcommand{\be}{\begin{equation}}
\newcommand{\ee}{\end{equation}}
\newcommand{\bdm}{\begin{displaymath}}
\newcommand{\edm}{\end{displaymath}}
\newcommand{\bea}{\begin{eqnarray}}
\newcommand{\eea}{\end{eqnarray}}

\newcommand{\fs}{\; \; .}
\newcommand{\co}{\; \; ,}
\newcommand{\al}{&\!\!\!\!}

\newcommand{\lvac}{\langle 0|\,}
\newcommand{\rvac}{\,|0\rangle}

\newcommand{\AmS}{{\protect\the\textfont2
  A\kern-.1667em\lower.5ex\hbox{M}\kern-.125emS}}

\begin{document}
\title{Probing the quark condensate by means of
$\pi\pi$ scattering}

\author{
H. Leutwyler\address{\it Institut f\"{u}r theoretische Physik der
Universit\"{a}t Bern\\Sidlerstr. 5, CH-3012 Bern, SWITZERLAND}%
\thanks{Work supported in part by Swiss National Science Foundation}}

\maketitle

\begin{abstract}

The available experimental information is consistent with
the Weinberg predictions for the threshold parameters of $\pi\pi$-scattering.
The data, however, only
test those relations that are insensitive to
the quark masses $m_u$ and $m_d$.
Recent
theoretical progress leads
to remarkably sharp predictions for the behaviour of the S- and P-wave phase
shifts in the elastic region -- one of the very few low energy phenomena in
QCD, where theory is ahead of experiment and where new precision data would
subject the theory to a crucial test.

\end{abstract}

\vspace{1em}
\noindent
{\it Talk given at the} DA$\Phi$CE workshop, {\it Frascati, November 1996}

\section{Introduction}

My talk at the DA$\Phi$CE meeting consisted of two parts: an
introductory discussion of chiral perturbation theory and a report
concerning recent work on the application of this method to $\pi\pi$
scattering. In the following, I omit the first part, referring the reader to
one of the published reviews \cite{chpt}. In particular, the handbook
\cite{handbook} offers an excellent overview of the applications of interest
in connection with the experiments planned at DA$\Phi$NE. This reference
also describes the status of our current knowledge of low energy
$\pi\pi$-scattering, both from the phenomenological point of view
\cite{Morgan Pennington handbook} and
within the framework of chiral perturbation theory
\cite{Gasser handbook}, \cite{Knecht handbook}. For a comprehensive discussion
of the general properties of the $\pi\pi$ scattering amplitude,
see refs.\ \cite{Martin}, \cite{Wanders 1971}.

Since the mass difference $m_u-m_d$ is small, the strong interaction
approximately conserves isospin. Neglecting the electromagnetic
interaction, the various elastic reactions among two pions may
be represented by a single scattering amplitude
$A(s,t,u)$. Only two of the Mandelstam variables are
independent, $s+t+u=4\,M_\pi^2$ and, as a consequence of Bose statistics, the
amplitude is invariant under an interchange of $t$ and $u$.

Equivalently, the amplitude may be described in
terms of the corresponding partial waves $t^I_\ell(q)$, where $q$ is the
momentum in the centre-of-mass-system, $\ell=0,1,2,\ldots$ is the
angular momentum and the
isospin index $I$ takes the values $I=0,1,2$. Bose symmetry entails that
only even values of $\ell+ I$ occur.
The first two terms in the
threshold expansion of the partial waves,
\be \mbox{Re}\,t_\ell^I(q)=q^\ell\,(a_\ell^I+ q^2\,b_\ell^I+\ldots)\co\ee
are referred to as scattering length ($a_\ell^I$) and effective range
($b_\ell^I$).

The pions are the Goldstone bosons of a spontaneously
broken approximate symmetry.
As shown by Weinberg \cite{Weinberg 1966}, chiral symmetry fully
determines the leading term in the low energy expansion of the $\pi\pi$
scattering amplitude,
in terms of the pion decay constant:
\be\label{current algebra} A(s,t,u)=\frac{s-M_\pi^2}{F_\pi^2}+O(p^4)\co\ee
The formula implies that (i) $\mbox{Re}\, A(s,t,u)$ contains
an Adler zero in the vicinity of $s=M_\pi^2$,
(ii) the scattering amplitude rapidly grows with the energy and (iii) in the
channels with $I=0$ and $I=1$, the interaction is attractive, while
for $I=2$ it is repulsive.

The present paper concerns the behaviour of the scattering amplitude in the
region $4\,M_\pi^2<s<4\,M_K^2$, where the Adler zero and the
pole due to $\rho$-exchange are the main features, which
determine the properties of the amplitude to a large extent \cite{Pennington
Portoles}. The $\rho$ may be incorporated into effective field theory models
by introducing a corresponding field.
Indeed, the tree graphs of a suitable model of this type provide a simple,
successful parametrization of the scattering amplitude
\cite{Schechter}.
The following discussion does not deal with the main features, however, but
addresses
the symmetry breaking effects due to the quark masses $m_u,m_d$. These
are very small and cannot reliably be analyzed on the basis of
such simplified models. Instead, I am using the Roy equations, which
follow from general principles and are perfectly suited for the low energy
analysis.

\section{Roy equations}
As was shown by Roy \cite{Roy}, analyticity and crossing symmetry imply that
the partial wave amplitudes satisfy a system of coupled dispersion relations
of the form \be \mbox{Re}\,t_\ell^I(q)=c_\ell^I(q)+
\sum_{I'=0}^2\sum_{\ell'}^\infty \int_4^\infty dq'\,Q_{\ell\ell'}^{I
I'}(q,q')\,\mbox{Im} \, t_{\ell'}^{I'}(q')\co\ee
The kernels $Q_{\ell\ell'}^{II'}(q,q')$ are explicitly
known kinematic quantities that may be expressed in terms of
associated Legendre functions and $c_\ell^I(q)$ represents a subtraction
polynomial.
In principle, a single subtraction constant suffices for the
above dispersion relations to converge, but the convergence is then rather
slow, so that the behaviour of the imaginary parts at high
energies play a significant role.
For the purpose of the low energy precision analysis which I wish to
discuss, it is
more appropriate to work with two subtraction constants, which may be
identified
with the S-wave scattering lengths $a_0^0,\,a_0^2$. The subtraction polynomials
then take the form \cite{Roy}
\bea
c_0^0(q)\al=\al a_0^0+\mbox{$\frac{1}{3}$}(2a_0^0-5a_0^2)q^2\co\hspace{2em}
     c_0^2(q)= a_0^2-\mbox{$\frac{1}{6}$}(2a_0^0-5a_0^2)q^2\co\\
c_1^1(q)\al=\al\mbox{$\frac{1}{18}$}
(2a_0^0-5a_0^2)q^2\co\hspace{3.85em}
     c_\ell^I(q)= 0\;\;\mbox{for}\;\;\ell\geq 2\fs\eea

The Roy equations express the real part of
$t_\ell^I(q)$ in
terms of the imaginary part.
We may represent the partial waves in terms of
the phase shift $\delta_\ell^I(q)$ and the elasticity
$\eta_\ell^I(q)$:
\be t_\ell^I(q)=\frac{\sqrt{M_\pi^2+q^2}}{2\,i\,q}\;\left\{\eta_\ell^I(q)\,
e^{\,2\,i\,\delta_\ell^I(q)}-1\right\}\fs\ee
Unitarity implies that in the elastic region $\eta_\ell^I(q)$ is equal to 1,
so that the real and imaginary parts of $t_\ell^I(q)$ are determined by a
single real function. For $s>16\,M_\pi^2$, on the other hand, unitarity
only leads to the bound $0\leq\eta_\ell^I(q)\leq 1$. Hence two real
functions are required to specify the behaviour of the partial waves there.
If the subtraction constants $a_0^0,\,a_0^2$ and the elasticities are taken as
known, the Roy equations amount to a set of coupled integral
equations for the phase shifts, which may be solved iteratively. It
is clear, however, that we cannot extract two real quantities from one
real equation. When using the Roy
equations to determine the partial waves, we thus need two categories of
input:
\begin{enumerate}
\item data in the inelastic region $\sqrt{s}>4\,M_\pi\simeq 550\,\mbox{MeV}$
\item values of the subtraction constants $a_0^0,\,a_0^2$
\end{enumerate}
The early literature on the problem
\cite{MMS}, \cite{Pennington}, \cite{Basdevant}
shows that the subtraction constants represent the
main source of uncertainty here -- the data available for
$\sqrt{s}>550\,\mbox{MeV}$ determine the values of the dispersion
integrals in the region below this point to within rather narrow
limits. The present experimental situation is described in
\cite{Morgan Pennington handbook}. For a thorough recent analysis of the data
in the range $600\,\mbox{MeV}<\sqrt{s}<1900\,\mbox{MeV}$, see
\cite{Bugg}.

The available experimental information does not allow us to determine the
S-wave scattering lengths very accurately. Various results are quoted in the
literature, for instance \cite{Nagels}
\be\label{phen}a_0^0=0.26\pm 0.05\,,\hspace{1em}a_0^2=-0.028\pm
0.012\,,\hspace{1em} 2\,a_0^0-5\,a_0^2=0.66\pm 0.05\,.\ee
The value for $a_0^0$ mainly relies on the
$K_{e_4}$-data of the Geneva-Saclay collaboration \cite{Rosselet} (the final
state interaction among the two pions allows a measurement of the phase
difference $\delta_0^0-\delta_1^1$ in the region $\sqrt{s}<M_K$).
The value inferred for $a_0^2$ then emerges from the correlation between the
two S-wave scattering lengths that was first noted by
Morgan and Shaw \cite{Morgan Shaw} ("universal curve"). The correlation is
due to the fact that, as mentioned earlier, one of the two subtraction
constants in the Roy equations is superfluous: The
combination $2\,a_0^0-5\,a_0^2$ may be represented as a convergent dispersion
integral over the imaginary part of the amplitude. The number given above
indicates that the available experimental information about the imaginary part
allows us to evaluate the dispersion integral to within about
10 \%.

\section{Wanders sum rules}
As an illustration of the statement that, if the two S-wave scattering lengths
are
taken as known, the available data allow a remarkably precise determination of
the scattering amplitude in the low energy region, I briefly discuss
one of the Wanders sum rules \cite{Wanders 1966}. Evaluating the second
derivative
of the Roy equation for $t_0^0(q)$ at $q=0$, we obtain a sum rule for the
effective range:
\be b_0^0=\frac{1}{3}(2\,a_0^0-5\,a_0^2)+\mbox{dispersion integral} \fs\ee
The relation represents a variant of the sum rule underlying the universal
curve. The main difference is that the dispersion integral occurring here
converges more rapidly, so that the result is less sensitive to the
experimental uncertainties in the high energy region. Moreover, the
contribution from the scattering
lengths dominates the result -- the one from the dispersion integral
amounts to about 20 \%. If $a_0^0$ and $a_0^2$ are assumed known,
the sum rule determines the
effective range to within a few percent. Similar sum rules exist for the
other threshold parameters,
$a_1^1,\,b_1^1,\,a_2^0,\,a_2^2,\,\ldots\,$ \cite{ATW}, \cite{Stern
pipi}, \cite{AB}.

Note that the threshold region generates significant contributions to the
dispersion integrals, so that an iterative procedure must be used to obtain
reliable values.
The sum rules for the threshold parameters may be viewed as
limiting cases of the Roy equations. A more systematic analysis of the
low energy behaviour of the partial
waves calls for an iterative solution of these equations. The main point here
is that $a_0^0,\,a_0^2$ represent the essential low energy parameters.
Once these are known, the available data suffice to pin down
the phase shifts quite accurately, despite the fact that, in the
elastic region, the experimental error bars are large.

\section{Chiral perturbation theory}
The representation (\ref{current algebra}) implies that the
expansion of the S-wave
scattering lengths in powers of the quark masses $m_u,\,m_d$ starts with
\cite{Weinberg 1966}
\be a_0^0=\frac{7\,M_\pi^2}{32\,\pi\,F_\pi^2}+O(M_\pi^4)\co \hspace{3em}
     a_0^2=-\frac{M_\pi^2}{16\,\pi\,F_\pi^2}+O(M_\pi^4)\fs
\ee
These expressions are proportional to $M_\pi^2$, i.e.\ the S-wave scattering
lengths vanish in the chiral limit, $m_u,\,m_d\rightarrow 0$. Indeed, chiral
symmetry prevents Goldstone bosons of zero momentum from interacting with one
another, so that, in the symmetry limit, the effective Lagrangian
exclusively contains
derivative couplings. The S-wave scattering
lengths therefore offer direct
probes of the chiral symmetry breaking generated by the quark masses. Since
$m_u$ and $m_d$ are very small, these effects
are difficult to measure. In the case of the $I=0$ S-wave, for
instance, where the threshold expansion starts with
$\mbox{Re}\,t_0^0(q)=a_0^0+q^2\,b_0^0+\ldots\,$, the scattering
length only dominates in the immediate vicinity of threshold: For
$q>M_\pi$, the effective range term is more important,
because this term does not represent a symmetry breaking effect (the low
energy expansion starts with
$b_0^0=1/(4\,\pi\,F_\pi^2)+\ldots\,$).

Weinberg's low energy theorem only accounts for the leading term in the
expansion of the scattering lengths in powers of $m_u,m_d$. The
higher order corrections may be worked out by means of chiral perturbation
theory, which yields a series of the form
\be\label{epsilon}
a_0^0=\frac{7\,M_\pi^2}{32\,\pi\,F_\pi^2}\{1+\epsilon_1+\epsilon_2+\ldots\}\co
\hspace{2em}\epsilon_n=O(M_\pi^{2n})\fs\ee
The explicit expression for the one loop correction reads \cite{GL 1983}
\be \epsilon_1=\frac{5\,M_\pi^2}{84\,\pi^2\,F_\pi^2}
(\bar{l}_1+2\,\bar{l}_2-\mbox{$\frac{3}{8}$}\,\bar{l}_3 +\mbox{$\frac{21}{10}$}
\,\bar{l}_4+\mbox{$\frac{21}{8}$})\co\ee
where $\bar{l}_1,\ldots\,,\,\bar{l}_4$ are effective coupling
constants of the chiral Lagrangian, normalized at running scale $\mu=M_\pi$.

\section{Effective coupling constants}
The coupling constants $\bar{l}_1$ and $\bar{l}_2$ may be determined in several
independent ways. In particular, an analysis of the $K_{e_4}$-data
yields \cite{BCEG handbook}
$\,\bar{l}_1=-1.7\pm 1.0$, $\bar{l}_2=6.1\pm
0.5$. Alternatively, the values of these coupling constants may be
determined by comparing the chiral
representation of the $\pi\pi$ scattering amplitude with the data
\cite{GL 1983}, \cite{Donoghue Ramirez Valencia pipi}. Recent work in this
direction
\cite{ATW}, \cite{Stern pipi}, \cite{AB} has clarified the role of the higher
order
contributions considerably. It leads to a coherent low energy representation
of the
scattering amplitude that confirms the results obtained from $K_{e_4}$-decay.
Moreover,
the physics underneath
the phenomenological values of these couplings is well understood: The main
contribution stems from the singularity generated by the exchange of a
$\rho$-meson \cite{Appendix Annals}, \cite{Donoghue Ramirez Valencia
resonances}, \cite{Pennington Portoles}, while
the exchange of particles of spin 0 or 2 only gives rise to comparatively small
corrections \cite{EGPdeR}, \cite{Toublan}.

The coupling constant $\bar{l}_4$ may be determined on the basis of
SU(3)$\times$SU(3), using the observed value of the ratio $F_K/F_\pi$ of decay
constants, or, alternatively, the slope of the scalar form factor occurring in
$K_{\ell_3}$-decay \cite{GL 1983}. The result, $\bar{l}_4=4.3\pm 0.9$,
predicts a value for the combination $2\,a_0^0-5\,a_0^2$ that is in good
agreement with the universal curve of Morgan and Shaw.
Moreover, there is an independent determination of $\bar{l}_4$, based on a
dispersive analysis of the scalar pion form factor \cite{DGL}. The value
obtained in that reference for the scalar "charge" radius of the pion,
$\langle r^2\rangle_\pi^s= 0.61\, \mbox{fm}^2$, corresponds to
$\bar{l}_4=4.6$ and thus also confirms the above estimate.

Concerning $\bar{l}_3$, however, the direct experimental
information is very meagre.
The physical significance of this coupling
constant is best seen in the formula for the mass of the pion. The
Gell-Mann-Oakes-Renner relation,
\be F_\pi^2\,M_\pi^2=(m_u+m_d)\,|\langle 0|\bar{u}u|0\rangle |+O(m^2)\co\ee
states that, at leading
order of the expansion in powers of $m_u$ and $m_d$, the square of the
pion mass is linear in the quark masses.
The constant $\bar{l}_3$ is the coefficient of the correction occurring at
second order:
\bea M_\pi^2\al= \al M^2
-\frac{\bar{l}_3\,M^4}{32\,\pi^2\,F_\pi^2}+O(M^6)\co\\
M^2\al\equiv\al (m_u+m_d)\, B_0 \co\hspace{4em}
B_0=\mbox{lim}
\rule[-0.7em]{0em}{0em}_{\hspace{-1.6em}m\rightarrow 0}\;\frac{1}{F_\pi^2}
\,|\langle 0|\bar{u}u|0\rangle |\fs\nonumber\eea

In principle, the dependence of the pion mass on
$m_u$ and $m_d$ can be determined on the lattice \cite{lattice}. The available
lattice results
do indicate that $M_\pi^2$ grows linearly with the quark mass, but it is
difficult to estimate the uncertainties arising from the fact
that, for the time being, dynamical quarks with realistic masses are out of
reach.

In the standard picture, the correction term is expected to be very small.
In particular, none of the singularities generated by low lying resonances
yields a large contribution to $\bar{l}_3$. Crude estimates, for instance
the one based on
$\eta'$-dominance for $L_7$ and on the Zweig rule, lead to
$\bar{l}_3\simeq 3$, indicating that the departure from the linear mass
formula is of order $2\,\%$. The expression (\ref{epsilon}) shows
that in the case of the scattering length, the contribution due to $\bar{l}_3$
is smaller than the one in the mass formula by the factor $\frac{5}{7}$.
Unless the theoretical estimates for $\bar{l}_3$ are entirely incorrect, the
precise value of this constant does not significantly affect the
prediction for $a_0^0$.

\section{Value of $a_0^0$}
Weinberg's formula leads to $a_0^0=0.16$.
With the above estimates for the coupling constants, the corrections increase
this number to 0.20, lower than the experimental value
(\ref{phen}) by about one standard deviation.

At first sight, the corrections appear to be unreasonably large.
After all, these are of the type
$\epsilon_1=(m_u+m_d)/\Lambda_0$.
A familiar rule of thumb states that the scale $\Lambda_0$ should be
of order 500 MeV or 1 GeV, indicating that the corrections should be of the
order of a few percent.
The same bookkeeping also applies within chiral perturbation theory, where
the correction is given by an expression of the
type $\epsilon_1= c\,M_\pi^2/(4\pi F_\pi)^2$ and $M_\pi^2/(4\pi F_\pi)^2=0.014$
is
a very small number.

The resolution of the paradox is that the coefficient
contains an infrared singularity. The explicit expression is of the form
$c=9\;\mbox{ln}(\Lambda_1/M_\pi)$, where
$\Lambda_1$ is independent of the quark masses. In the limit
$m_u,m_d\rightarrow 0$, the coefficient $c$ therefore diverges
logarithmically. Infrared
singularities of this type are a commom occurrence in chiral perturbation
theory. In the present case, the
logarithm has an unusually large coefficient, partly because we are
considering the value of the partial wave amplitude at
threshold, i.e. at the branch point singularity associated with two-pion
intermediate states.
The estimates for the coupling constants given above yield
$c\simeq 20$, which corresponds $\Lambda_1\simeq 1.3\,\mbox{GeV}$, a perfectly
reasonable scale.

In the meantime,
the $\pi\pi$ scattering amplitude has been worked out to two loops of chiral
perturbation theory \cite{BCEGS}. Using theoretical estimates for the
additional
coupling constants occurring at that order,
the authors obtain $a_0^0=0.217$.
The infrared singularities again yield the
dominating contribution.
A systematic analysis of the effective couplings
occurring in the two loop result is
under way \cite{phen two loop}. Two types of coupling constants occur: those
which survive in the chiral limit
and which manifest themselves in the momentum dependence of the amplitude and
those which account for the symmetry breaking due to $m_u,\,m_d$.
As demonstrated in ref.\
\cite{Stern pipi}, the former may be determined phenomenologically, on the
basis of the available data. In the standard picture, the latter
only generate very small effects, dominated by the
contribution
from $\bar{l}_4$. A rough estimate of the uncertainties in the various elements
of the calculation indicates that the two loop representation of the
scattering amplitude will allow us to predict
the value of $a_0^0$ to an accuracy of 2 or 3 \%.

It should be noted that, at this level of precision,
electromagnetic contributions are not negligible. In
particular, the above numbers rely on
$M_\pi=M_{\pi^+}$, the scale conventionally used to express dimensionful
quantities in pion mass units. If the electromagnetic
interaction is turned off, the mass of the charged pion decreases
by about 4 MeV. The correction reduces the prediction for the S-wave scattering
lengths by 6 \%. Also, the pion decay constant
must be corrected for radiative effects, which act in the opposite
direction. Taken together, these modifications reduce the two loop result
to $a_0^0=0.208$. For a discussion of isospin breaking
in the $\pi\pi$ scattering amplitude, I refer to \cite{em}.

\section{Need for low energy precision experiments}
As discussed above, the prediction for $a_0^0$ relies on theoretical estimates
for the coupling constant $\bar{l}_3$. Stern and
collaborators have emphasized that
the direct experimental information on the magnitude of this coupling
constant leaves much to be desired
\cite{Knecht handbook}, \cite{Stern pipi}.
Expressed in terms of $\bar{l}_3$, the
experimental uncertainty in the value of $a_0^0$ roughly corresponds to
\be a_0^0=0.26\pm 0.05\;\;\;\Longleftrightarrow\;\;\; \bar{l}_3=-60\pm 60\fs\ee
If it should turn out that the data
confirm the central value, the immediate conclusion to draw would be that
$\bar{l}_3$ is negative
and very large. This in turn would imply that the higher order "corrections"
which the Gell-Mann-Oakes-Renner relation neglects are as big as the "leading"
term -- the standard theoretical picture underlying our current
understanding of the low energy structure of QCD
would be proven wrong.
In particular, the standard pattern of the light quark masses, where
$(m_u+m_d)/m_s\simeq M_\pi^2/M_K^2$, would then be incorrect.

As discussed in detail in refs.\ \cite{Knecht handbook}, \cite{Stern pipi},
the standard picture
relies on the hypothesis that the quark condensate $\lvac \bar{u}u\rvac$ is the
leading order parameter of the spontanously broken symmetry.
The condensate plays a role analogous to the one of the
spontaneous magnetization of a magnet.
In that case, it is well-known that spontaneous symmetry breakdown
occurs in two quite different modes:
ferromagnets and antiferromagnets. For the former, the magnetization
develops a non-zero expectation value,
while for the latter, this does not
happen. In either case, the symmetry is spontaneously broken (for a
discussion of the phenomenon within the effective Lagrangian framework,
see \cite{Nonrel}). The example
illustrates that operators which are allowed by the symmetry to pick up
an expectation value may but need not do so. In particular, general principles
do not exclude the possibility that the quark condensate vanishes in the chiral
limit.

Since the form of the effective chiral Lagrangian is determined by
symmetry, the standard expression, characterized by the coupling constants
$F_\pi,\, B,\,\bar{l}_1,\,\bar{l}_2,\,\bar{l}_3,\,\ldots$ also applies if
$\bar{l}_3$ is taken to be large, but it then becomes inconsistent to treat
the interaction generated by this coupling constant as a perturbation.
Instead, the term proportional to $\bar{l}_3$ must then be included
among the leading contributions of order $p^2$: The effective Lagrangian
remains the same, but the chiral perturbation series must be reordered
("generalized chiral perturbation theory").
The implications for some of the low energy observables of interest, in
particular
also for $\pi\pi$ scattering have been worked out in detail \cite{Knecht
handbook}, \cite{Stern pipi}.

If the picture underlying the standard framework should turn out
to be incorrect, much of the predictive power of
the effective theory would be lost. In the case of $a_0^0$, for
instance, the generalized scenario does not yield a prediction, because
$\bar{l}_3$ is treated as a free parameter. Another example is the
Gell-Mann-Okubo formula for the masses of the pseudoscalar octet, which
represents a very neat check of the standard framework -- the
generalized scheme does not lead to such a formula, but can
accomodate the observed mass pattern if the corresponding effective coupling
constants are properly tuned.

Quite irrespective, however, of whether or not the alternative
is theoretically attractive, it is important to subject the issue to
experimental test.
Low energy precision measurements of the $\pi\pi$ scattering
amplitude would allow us to settle the matter. In particular, an
analysis of the momentum distribution in the decay $K\rightarrow\pi\pi
e\nu$ does
provide a test of those predictions that concern the breaking of chiral
symmetry due to $m_u$ and $m_d$. At DA$\Phi$NE, this can be done, with
significantly better statistics than in the Geneva-Saclay
experiment. Moreover, the properties of the transition matrix elements
are now under better control, so that the data analysis can be
performed on a more solid basis \cite{Baillargeon Franzini}.

There is a beautiful alternative proposal due to Nemenov
\cite{Nemenov}, based on the observation that
$\pi^+\pi^-$ atoms decay into a pair of neutral pions, through the
strong transition $\pi^+\pi^-\!\rightarrow\!\pi^0\pi^0$. Since the
momentum transfer nearly vanishes, the decay rate is proportional to
the square of the combination $a_0^0\!-\!a_0^2$ of
S-wave $\pi\pi$ scattering lengths.
A measurement of the
lifetime of a $\pi^+\pi^-$ atom would thus also
allow us to decide whether or not the quark condensate represents the
leading order parameter. An experiment with this goal is under way at CERN.

\end{document}